# A Self-Consistent Dynamical Model for the *COBE* Observed Galactic Bar and Its Application to Microlensing

HongSheng Zhao

*Max-Planck-Institute für Astrophysik, 85740 Garching, Germany*
*Email: hsz@MPA-Garching.MPG.DE*

**Abstract.** A self-consistent stellar dynamical model for the Galactic bar is constructed from about 500 numerically computed orbits with an extension of the Schwarzschild technique. The model fits the *COBE* found asymmetric boxy light distribution and the observed stellar kinematics of the bulge. The model potential is also consistent with the non-circular motions of the HI and CO velocity maps of the inner Galaxy. We also use the stellar bar model to construct a N-body model to study stability and a microlensing map towards the bulge, which can account for the observed optical depth and the event duration by the MACHO and OGLE collaborations. The technique used here can be applied to interpret light and velocity data of external bulges/bars and galactic nuclei.

## 1. Introduction

It is now widely known that our Galaxy has a central bar with its near end on the positive Galactic longitude side (see the review by Kuijken in this proceeding). Two strong evidences for this picture come from the asymmetric light distribution found by the *COBE* team (Weiland et al. 1994) and the non-circular motion of the HI and CO clouds (Binney et al. 1991). Interestingly the large microlensing optical depth towards the bulge found by the OGLE and MACHO teams (Alcock et al. 1995, Udalski et al. 1994) is also in agreement with most lenses being in the near side of a massive bar in the center, pointing nearly towards us (Zhao et al. 1995). While the observations can directly rule out existing oblate rotator models (Kent 1992), it is much harder to device a bar model that fits the observations qualitatively. The traditional N-body approach to make a bar from an unstable disk is often not good enough for quantitative interpretations although some remarkable progress has been made in this direction (see Fux et al. in this proceeding).

In this talk, I show an equilibrium model for the stellar bar that is made particularly to fit the *COBE* light distribution and the stellar kinematics of the Galactic bar. The basic technique is as Schwarzschild (1979), but we have implemented many necessary technical modifications (see Zhao 1994 PhD thesis and Zhao 1995). The model is made by populating 500 orbits in a fixed bar potential. The mass on each orbit is determined to fit the observations of the light and kinematics in the least square sense (Figure 1) and also to reproduce the potential self-consistently. In particular, the model fits the asymmetric boxy shape of the observed light distribution (the dust-subtracted K band *COBE* map from Weiland et al. 1994 is the dotted line, and the reprojection of the volume density model from Dwek et al. 1995 is the dashed line. the model is in solid line) and the fall-off of the observed radial velocity dispersions on the minor axis (data from Terndrup et al. 1995, Sharples et al. 1990), and the proper motion dispersions at Baade's Window (from Spaenhauer et al. 1992) and the vertex deviation $\sigma_{lr}$ (from Zhao et al. 1994). Note a (probably too) simple Miyamoto-Nagai disk has been added to the bar's surface density map for direct comparison with the *COBE* map.

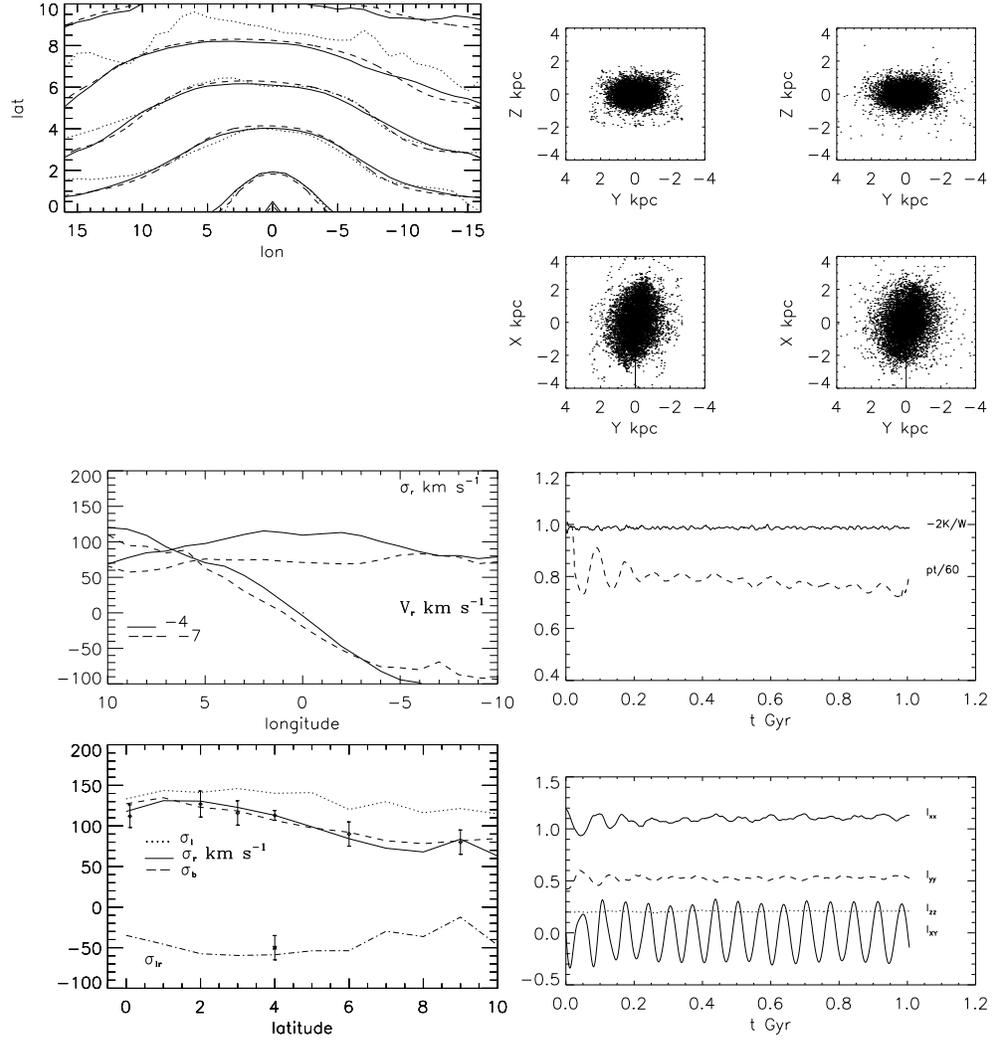

Figure 1. The left and right panels show the self-consistent model from the Schwarzschild technique and the evolution of its N-body counterpart respectively. The left three panels are the predicted surface density map (solid contours, 1 magnitude spacing in between), the kinematics along $b = -4^\circ$ and $b = -7^\circ$, and the radial dispersion ($\sigma_r$), the proper motions ($\sigma_l$, $\sigma_b$) and the cross term $\sigma_{lr}$ along the minor axis. The corresponding observations are also plotted. The right panels show the N-body realization in the beginning (to the left), and the configuration after evolving for 10 rotation periods (to the right) both in the face-on view and edge-on view. The lower two panels shows the global equilibrium indicators as functions of time more quantitatively, which include the Virial equilibrium indicator $-2K/W$ and the normalized pattern speed (both should be unity if in steady state), the three moments of inertia $I_{xx}$, $I_{yy}$ and $I_{zz}$ along the three principal axes (should all be constant) and the rest frame cross term $I_{XY}$ (should be sinusoidal).

The stability of the model is tested by first converting the orbit model to an N-body model. The conversion is done by spreading 50K particles randomly in the phase of each orbit with a number in proportion to the weight of each orbit. The evolution of the N-body bar is followed with the Self-Consistent Field method code (Hernquist and Ostriker 1992) for a Gigayear. In the case with a rigid disk potential, the bar is stable for at least 1 Gyr with only small evolution of the pattern speed and the shape (Figure 1). Note the elongated bar shape in the face-on view and the boxyness in the edge-on view are similar at two epochs. The final bar has settled down to dynamical equilibrium. This is interesting since the bar model also has a self-consistent central cusp ($\rho \sim r^{-1.85}$) with a mass 5-10% of the bar. We find that the central cusp does not cause the bar to dissolve rapidly through scattering the boxy orbits.

The self-consistent model can directly tell us how different orbit families are populated. The model bar's mass is divided between explicitly integrated direct orbits and some orbits (which I call collective-orbits) with an implicit isotropic distribution function of $f = f(E_J)$. The advantage of using this hybrid representation of the bar's phase space is to get around integrating the ill-understood chaotic orbits explicitly without missing any necessary orbit families of the bar. We find that although the dominant orbits in the model are still the direct boxy orbits (60% in mass), which are responsible for both the boxy contours in the *COBE* map and the rapid rotation of the bulge, the rest of the mass is in the collective-orbits, which implicitly contain 2/3 chaotic orbits and 1/3 retrograde orbits. As one does not expect significant amount of retrograde or chaotic orbits be populated during the formation of the bar from the disk, we argue that the large fraction seen in the model poses a possible challenge to this canonical scenario.

As the stars and the gas share the same potential, it is of interest to study the response of gas in the *COBE* potential. Binney et al. (1991) argue that the pressure force and dissipational collisions tend to beam gas clouds on non-self-intersecting closed orbits, and in particular, the shape of the non-self-intersecting $x_1$ orbits should to zeroth order match the dynamical boundaries of the HI and CO gas clouds in the longitude-velocity plane. We have reexamined this interpretation but now with a realistic potential based on fitting stellar observations. We find a fast bar with pattern speed 60 km/s/kpc and a bar angle of $10-20$ degrees has a vertical edge at $l = 2^o$ due to the nearly cusped $x_1$ orbits (the vertical loop), which could fit the parallelogram of CO (not shown, but similar to Binney et al. Figure 2). The terminal velocities (the crosses) of the non-cusped $x_1$ orbits has a rapid rise and fall off which matchs the terminal velocity of the HI map (Figure 2). Models with bigger bar angles and/or smaller pattern speed appear to be worse in the fit. The HI map was kindly made electronicly available by Harvey Liszt.

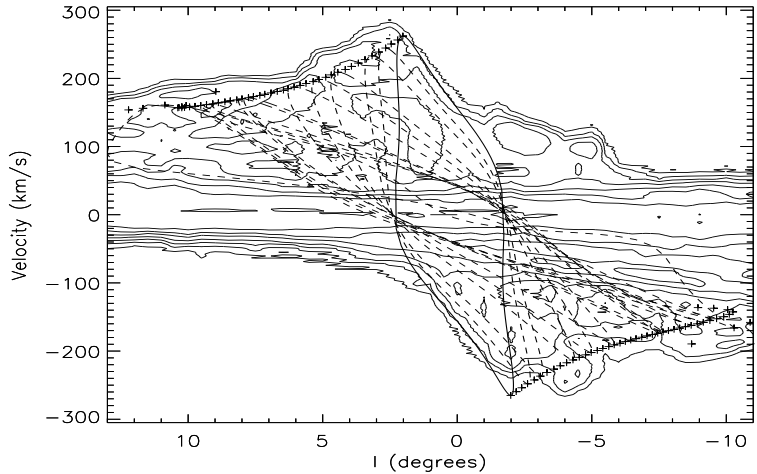

Figure 2.   overplots the HI map with the $l-v$ digram of the $x_1$ orbits for our best model.

A microlensing map of the bar is another by-product of the model. It is built by flagging the particles in the converted N-body model as lenses or sources and "observing" them along different line-of-sights. We find that the optical depth of the model at MACHO and OGLE observed fields is a few times $10^{-6}$. The typical event should last about 30 days if the lenses were all one solar. Comparing these values with the observed 20 day time scale (Alcock et al. 1995, Udalski et al. 1994), we estimate that the average mass of the lenses is $0.4 M_\odot$, well above the brown dwarf limit of $0.1 M_\odot$ (other results are summarized in Zhao et al. 1995a,b). We argue that most of the lens seen in these experiments are luminous stars, which have the hope to be detected with other methods as well as with microlensing.

In summary, we have built a sophisticated model for the Galactic bar that is consistent with a variety of recent observations. The model is likely to be stable and unique. More rigorous tests of stability (simulations with a live disk and halo) and exploring the full range of plausible models still remain to be done. The model results will be made electronically available to the community in the near future.

**Acknowledgments.**   This work is a bulk part of the PhD thesis work done in Columbia University. I thank David Spergel and Michael Rich for advices and supports during this work.

### Discussion

*D. Pfenniger*: How do you choose your initial conditions for the N-body run from the Schwarzschild model?

*Zhao*: I select the particles so that the number of particles on each orbit is proportional to the weight on that orbit derived from the Schwarzschild method. Inside each orbit, particles are randomly distributed in phase. So the N-body run is based on a random realization of the bar orbits.

*Ben Werner*: The match of the $x_1$ orbits to the tangent point profile in the HI $l - v$ diagram is impressive (and good because it does away with the dip in the Clemens et al. rotation curve), but so far it doesn't get enough material into the forbidden regions (I think that is because the bar is so close to end-on in your model.)

*Zhao*: It is true that the $x_1$ orbits in the best model do not fully occupy the forbidden region. But I suspect that is because full hydrodynamics of the gas has not been included in the potential model.